\documentclass[twocolumn,showpacs]{revtex4}
\usepackage{graphicx}

\begin{document}
\title{Conservation laws for dynamical black holes}
\author{Sean A. Hayward}
\affiliation{Center for Astrophysics, Shanghai Normal University, 100 Guilin 
Road, Shanghai 200234, China}
\date{Revised 4th September 2006}

\begin{abstract}
An essentially complete new paradigm for dynamical black holes in terms of 
trapping horizons is presented, including dynamical versions of the physical 
quantities and laws which were considered important in the classical paradigm 
for black holes in terms of Killing or event horizons. Three state functions 
are identified as surface integrals over marginal surfaces: irreducible mass, 
angular momentum and charge. There are three corresponding conservation laws, 
expressing the rate of change of the state function in terms of flux 
integrals, or equivalently as divergence laws for associated conserved 
currents. The currents of energy and angular momentum include the matter 
energy tensor in a physically appropriate way, plus terms attributable to an 
effective energy tensor for gravitational radiation. Four other state 
functions are derived: an effective energy, surface gravity, angular speed and 
electric potential. There follows a dynamical version of the so-called first 
law of black-hole mechanics. A corresponding zeroth law holds for null 
trapping horizons.
\end{abstract}
 \maketitle

{\em Introduction.} Black holes are now generally regarded as astrophysical 
realities, which are expected to be major sources of gravitational waves, 
prompting extensive studies of dynamical, strong-field processes such as 
binary mergers. The textbook theory of black holes, however, mostly concerns 
stationary black holes or physically unlocatable event horizons 
\cite{BCH,HE,MTW,Wal}. In recent years, a new paradigm for dynamical black 
holes has been developed in terms of trapping horizons 
\cite{bhd,1st,AK,BF,en,Gou,j15,bhd5}, hypersurfaces where light is momentarily 
caught by the gravitational field, which locate the black hole in a practical 
way. For physical reasons, unique measures of mass and angular momentum for 
such locally defined black holes are desired, together with conservation laws 
describing how they change in terms of the fluxes of energy and angular 
momentum of the infalling matter and gravitational radiation, so as to 
describe how a black hole grows and spins up or down. 

This Letter reports a generically unique definition of angular momentum, 
obtained directly from the Komar integral \cite{Kom}, satisfying a 
conservation law with a similar form to the energy conservation law \cite{en}. 
Adding charge conservation for generality, this allows general definitions of 
all the key physical quantities of the classical paradigm, plus dynamical 
versions of the so-called first and zeroth laws of black-hole mechanics 
\cite{BCH}. A preliminary report was given previously \cite{j15} and a more 
detailed description is given in a longer article \cite{bhd5}.

\begin{figure}
\includegraphics[height=15mm]{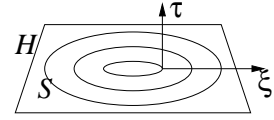}
\caption{A hypersurface $H$ foliated by spatial surfaces $S$, with generating 
vector $\xi$ and its normal dual $\tau=\xi^*$.} \label{normal}
\end{figure}

{\em Geometry.} General Relativity will be assumed, with space-time metric 
$g$. A one-parameter family of spatial surfaces $S$ locally generates a 
foliated hypersurface $H$. Labelling the surfaces by a coordinate $x$, they 
are generated by a vector $\xi=\partial/\partial x$, which can be taken to be 
normal to the surfaces, $\bot\xi=0$, where $\bot$ denotes projection onto $S$. 
A duality operation on normal vectors $\eta$, $\bot\eta=0$, yields a dual 
normal vector $\eta^*$ defined by $\bot\eta^*=0$, $g(\eta^*,\eta)=0$, 
$g(\eta^*,\eta^*)=-g(\eta,\eta)$, and in particular $\tau=\xi^*$ is  normal to 
$H$ (Fig.\ref{normal}). The coordinate freedom is $x\mapsto\tilde x(x)$ and 
choice of angular coordinates on $S$, under which all the key formulas will be 
invariant. 

The expansion $\theta_\eta=L_\eta\log(d^2\!A)$ along a normal vector $\eta$, 
where $L$ denotes the Lie derivative and $d^2\!A$ the area form of $S$, can be 
expressed in terms of the expansion 1-form $\theta=d\log(d^2\!A)$ as 
$\theta_\eta=\theta(\eta)$. It is convenient to use two future-pointing null 
normal vectors $l_\pm$ to $S$, $g(l_\pm,l_\pm)=0$, $\bot l_\pm=0$, since their 
directions are unique. Then a normal vector $\eta$ has components $\eta^\pm$ 
along $l_\pm$, in particular $\xi=\xi^+l_++\xi^-l_-$, $\tau=\xi^+l_+-\xi^-l_-$ 
and $g^{-1}(\theta)=-e^f(\theta_-l_++\theta_+l_-)$, where $f$ is a 
normalization function defined by $e^{-f}=-g(l_+,l_-)$ and 
$\theta_\pm=\theta(l_\pm)$ are the null expansions. One may adapt $l_\pm$ to 
$H$ via $l_A(dx^B)=\delta_A^B$, where $x^\pm$ are coordinates labelling the 
null hypersurfaces generated from $S$ in the normal directions 
\cite{bhd,1st,en,bhd5}.
 
A {\em trapping horizon} \cite{bhd,1st,en} is a hypersurface $H$ foliated by 
marginal surfaces, where $S$ is {\em marginal} if one of the null expansions, 
$\theta_+$ or $\theta_-$, vanishes everywhere on $S$. A confusing multiplicity 
of names have been proposed for trapping horizons under various extra 
conditions, but such conditions will be largely irrelevant here, since all the 
equations and results, except where specifically noted, hold for any trapping 
horizon with compact $S$.

{\em Angular momentum.} The standard definition of angular momentum for an 
axial Killing vector $\psi$ and at spatial infinity is the Komar integral 
\cite{Kom}
\begin{equation}
J[\psi]=-\frac1{16\pi}\oint_S\epsilon_{\alpha\beta}\nabla^\alpha\psi^\beta 
d^2\!A \label{komar}
\end{equation}
where $\epsilon$ is the binormal to $S$ and Newton's constant is set to unity. 
Here $(\alpha,\beta\ldots)$ denote general indices, $(a,b\ldots)$ will denote 
transverse indices and $(A,B\ldots)=\pm$ will denote normal indices, e.g.\ 
$\epsilon^{AB}=e^f(l_-^Al_+^B-l_+^Al_-^B)$.

\begin{figure}
\includegraphics[height=3cm]{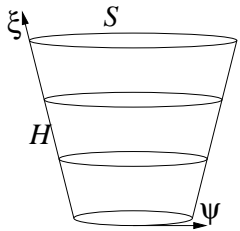}
\caption{A transverse vector $\psi$.} \label{transverse}
\end{figure}

Now consider $\psi$ to be a general transverse vector, $\bot\psi=\psi$ 
(Fig.\ref{transverse}). Since $\epsilon_{\alpha\beta}\psi^\beta=0$, the Komar 
integral can be rewritten as 
\begin{equation}
J[\psi]=\frac1{8\pi}\oint_S\psi^a\omega_ad^2\!A \label{am}
\end{equation}
where the twist \cite{dne} $\omega_a=\frac12e^fh_{a\beta}[l_-,l_+]^\beta$ is a 
transverse 1-form, $\bot\omega=\omega$, and $h$ is the induced metric on $S$. 
The twist encodes the non-integrability of the normal space, thereby providing 
a geometrical measure of rotational frame-dragging. It is an invariant of a 
non-null foliated hypersurface $H$, so $J[\psi]$ is also an invariant. It can 
be checked \cite{bhd5} to recover the standard definition of angular momentum 
for a weak-field metric \cite{MTW}, with $\omega$ determining the precessional 
angular velocity of a gyroscope due to the Lense-Thirring effect.

There are several similar definitions of angular momentum, as clarified by 
Gourgoulhon \cite{Gou} and in the longer article \cite{bhd5}. Ashtekar \& 
Krishnan \cite{AK} gave a definition for dynamical horizons, involving a 
1-form which coincides with $\omega$ in that case. Ashtekar et al.\ \cite{IH} 
earlier gave a definition for isolated horizons, involving a 1-form which does 
not generally coincide with $\omega$. However, it can be made to coincide if 
the dual-null gauge is fixed in a natural way \cite{bhd5}. An earlier 
definition of angular momentum by Brown \& York \cite{BY} involves a 1-form 
which does not generally coincide with $\omega$, but does so if re-interpreted 
as adapted to the horizon. 

In all cases, there remains the question of choosing $\psi$ with properties 
appropriate to an axial vector. Ashtekar \& Krishnan \cite{AK} proposed that 
$\psi$ has vanishing transverse divergence, $D_a\psi^a\cong0$, where $D$ is 
the covariant derivative of $h$ and $\cong$ denotes equality on $H$. This 
condition holds if $\psi$ is an axial Killing vector, and can be understood as 
a weaker condition, equivalent to $\psi$ generating a symmetry of the area 
form rather than of the whole metric, since $L_\psi(d^2\!A)=D_a\psi^ad^2\!A$. 
Alternatively, assuming that the integral curves of $\psi$ are closed, as 
expected for an axial vector, it can be satisfied by choice of scaling of 
$\psi$, as discussed by Booth \& Fairhurst \cite{BF}. 

\begin{figure}
\includegraphics[height=25mm]{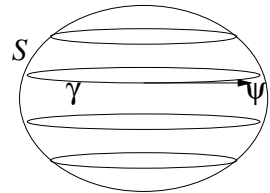}
\caption{Curves $\gamma$ of constant expansion $\theta_\xi$.} \label{axial}
\end{figure}

Spherical topology will be assumed henceforth for $S$, which follows from the 
topology law \cite{bhd} for outer trapping horizons, assuming the dominant 
energy condition. If there exist angular coordinates $(\vartheta,\varphi)$ on 
$S$ with $\psi=\partial/\partial\varphi$, completing coordinates 
$(x,\vartheta,\varphi)$ on $H$, then $L_\xi\psi\cong0$, as proposed by 
Gourgoulhon \cite{Gou}. Noting the commutator identity \cite{dne} 
$L_\xi(D_a\psi^a)-D_a(L_\xi\psi)^a=\psi^aD_a\theta_\xi$, assuming both 
conditions on $\psi$ forces $\psi^aD_a\theta_\xi\cong0$. This is automatic if 
$D\theta_\xi\cong0$, as in spherical symmetry or along a null trapping 
horizon. However, generically one expects $D\theta_\xi\not\cong0$ almost 
everywhere. It must vanish somewhere on a sphere, by the hairy ball theorem, 
but the simplest generic situation is that there are curves $\gamma$ of 
constant $\theta_\xi$ which form a smooth foliation of circles, covering the 
surface except for two poles (Fig.\ref{axial}). Assuming so, since $\psi$ is 
tangent to $\gamma$, one can find a unique $\psi$, up to sign, in terms of the 
unit tangent vector $\hat\psi$ and arc length $ds$ along $\gamma$: 
$\psi\cong\hat\psi\oint_\gamma ds/2\pi$, where the scaling ensures that the 
axial coordinate $\varphi$ is identified at 0 and $2\pi$. 

Then the angular momentum becomes unique up to sign, $J[\psi]=J$, the sign 
being naturally fixed by $J\ge0$ and continuity of $\psi$. This construction, 
if unique, will yield the axial Killing vector if one exists, in particular 
for a Kerr black hole \cite{j15,bhd5}. The definition can be applied in any 
situation where $D\theta_\xi\not\cong0$ almost everywhere, though the physical 
interpretation as angular momentum seems to be safest in the case of two 
poles, which locate the axis of rotation. Then $J$ is proposed to measure the 
angular momentum about that axis.

{\em Conservation of angular momentum.} Assuming the above conditions 
$L_\xi\psi\cong0$, $D_a\psi^a\cong0$, then
\begin{equation}
L_\xi J\cong-\oint_S(T_{aB}+\Theta_{aB})\psi^a\tau^Bd^2\!A \label{amc}
\end{equation}
holds along a trapping horizon, where $T$ is the matter energy tensor,
\begin{equation}
\Theta_{a\pm}=-{1\over{16\pi}}h^{cd}D_d\sigma_{\pm ac} \label{eet}
\end{equation}
is the transverse-normal block of an effective energy tensor for gravitational 
radiation, and $\sigma_{\pm ab}=\bot L_\pm h_{ab}-\theta_\pm h_{ab}$ are the 
null shears, which are transverse, $\bot\sigma_\pm=\sigma_\pm$, and traceless, 
$h^{ab}\sigma_{\pm ab}=0$. The proof is a calculation using the Einstein 
equations, the stated conditions and the Gauss divergence theorem \cite{bhd5}.

The null shears have previously been identified in the corresponding energy 
conservation law \cite{en} as encoding the ingoing and outgoing transverse 
gravitational radiation, via the energy densities 
$\Theta_{\pm\pm}=||\sigma_\pm||^2/32\pi$, which agree with expressions in 
other limits, such as the Bondi energy density at null infinity and the 
Isaacson energy density of high-frequency linearized gravitational waves. So 
the result implies that gravitational radiation with a transversely 
differential waveform will generally possess angular momentum density. In the 
absence of such terms, the conservation law is the standard surface-integral 
form of conservation of angular momentum, were $\psi$ an axial Killing vector, 
thereby describing the increase or decrease of angular momentum due to infall 
of co-rotating or counter-rotating matter respectively.

The identification of the transverse-normal block (\ref{eet}) of $\Theta$ 
appears to be new. Previous versions of angular momentum flux laws for 
dynamical black holes \cite{AK,BF,Gou} contain different terms, which are not 
in energy-tensor form, i.e.\ some tensor contracted with $\psi$ and $\tau$. 
They can be recovered by removing a transverse divergence from 
$\Theta_{aB}\psi^a\tau^B$, yielding 
$\sigma_\tau^{ab}D_b\psi_a/16\pi=\sigma_\tau^{ab}L_\psi h_{ab}/32\pi$, where 
$\sigma_\tau^{ab}=\tau^B\sigma_B^{ab}$ gives the shear along $\tau$. Such 
terms have been described by analogy with viscosity \cite{Gou}.

{\em Conservation of energy.} The recently derived energy conservation law 
\cite{en} will be stated here for comparison, modifying some notation. 
Introduce the area $A=\oint_Sd^2\!A$, the area radius $R=\sqrt{A/4\pi}$, the 
canonical time vector $k=(g^{-1}(dR))^*$ and the Hawking mass \cite{Haw}
\begin{equation}
M=\frac R2\left(1-\frac1{16\pi}\oint_S{*}g^{-1}(\theta,\theta)\right). 
\label{hm}
\end{equation}
Assuming the null energy condition, this is the irreducible mass $M\cong R/2$ 
of a future outer trapping horizon, $L_\xi M\ge0$, by the area law \cite{bhd}, 
which implies 
\begin{equation}
L_\xi A\ge0. \label{2nd}
\end{equation}
Then the energy conservation law \cite{en}
\begin{equation}
L_\xi M\cong\oint_S(T_{AB}+\Theta_{AB})k^A\tau^Bd^2\!A \label{ec}
\end{equation}
has a similar form to that of angular momentum (\ref{amc}). Of the ten 
conservation laws in flat-space physics, they are the two independent laws 
expected for an astrophysical black hole, which defines its own spin axis and 
centre-of-mass frame, in which its momentum vanishes.

It is appropriate to compare with other approaches, particularly that of 
Ashtekar \& Krishnan \cite{AK}, who derived a flux law for any transverse 
vector $\psi$. Partly this reflects a different viewpoint, that classes of 
flux laws were desired, different choices of the vectors $(k,\psi)$ yielding 
different quantities $(M,J)$. Here the aim has been to find unique, physically 
meaningful conserved quantities for a given black hole, respectively the 
irreducible mass $M$, which has a clear physical interpretation, and the 
angular momentum $J$ about the axis of rotation, which has been obtained 
generically from natural restrictions on $\psi$. Secondly, the conservation 
laws (\ref{amc}), (\ref{ec}) apply to any trapping horizon, whereas the 
Ashtekar-Krishnan formalism applies only to spatial trapping horizons, with 
null trapping horizons having been treated by the separate isolated-horizons 
formalism \cite{IH}, some connections having been drawn between the two, such 
as for slowly evolving horizons by Booth \& Fairhurst \cite{BF}. Null trapping 
horizons remain a degenerate case of the general framework, but there is a 
natural way to fix the additional gauge freedom \cite{bhd5}, consistently with 
weakly isolated horizons. Thirdly, the non-matter terms have here been 
identified as arising from an effective energy tensor $\Theta$ for 
gravitational radiation, which allows the physical interpretation as 
conservation laws rather than just flux laws. Fourthly, charge has not yet 
been included, except where angular momentum vanishes \cite{1st} or for 
isolated horizons \cite{IH}, as remedied below.

{\em Conservation of charge.} The surface-integral form of conservation of 
charge $Q$ is 
\begin{equation} L_\xi Q=-\oint_Sg(j,\tau)d^2\!A \label{cc}
\end{equation}
where the vector $j$ is the charge-current density. The conservation laws for 
energy (\ref{ec}) and angular momentum (\ref{amc}) take the same form 
\begin{equation} 
L_\xi M\cong-\oint_Sg(\tilde\jmath,\tau)d^2\!A, \quad L_\xi 
J\cong-\oint_Sg(\bar\jmath,\tau)d^2\!A \label{cl}
\end{equation}
by identifying current vectors $\tilde\jmath^B=-k_A(T^{AB}+\Theta^{AB})$, 
$\bar\jmath^B=\psi_a(T^{aB}+\Theta^{aB})$. 

It is noteworthy that the local differential form of charge conservation, 
$\nabla\!_\alpha j^\alpha=0$, generally does not hold for $\tilde\jmath$ or 
$\bar\jmath$. Instead one can obtain \cite{j15,bhd5}
\begin{equation}
\oint_S\nabla\!_\alpha\tilde\jmath^\alpha 
d^2\!A\cong\oint_S\nabla\!_\alpha\bar\jmath^\alpha d^2\!A\cong0. \label{qlc}
\end{equation}
This subtly confirms the view that energy and angular momentum in General 
Relativity cannot be localized \cite{MTW}, but might be quasi-localized, as 
surface integrals \cite{Pen}. The corresponding conservation laws have indeed 
been obtained in surface-integral but not local form.

{\em State space.} There are now three conserved quantities $(M,J,Q)$, forming 
a state space for dynamical black holes. Following various authors 
\cite{AK,BF,IH}, related quantities may then be defined by formulas satisfied 
by Kerr-Newman black holes, specifically those for the ADM energy 
\begin{equation} E\cong\frac{\sqrt{((2M)^2+Q^2)^2+(2J)^2}}{4M} \label{e}
\end{equation}
the surface gravity 
\begin{equation}
\kappa\cong\frac{(2M)^4-(2J)^2-Q^4}{2(2M)^3\sqrt{((2M)^2+Q^2)^2+(2J)^2}} 
\label{sg}
\end{equation}
the angular speed 
\begin{equation}
\Omega\cong\frac{J}{M\sqrt{((2M)^2+Q^2)^2+(2J)^2}} \label{as}
\end{equation}
and the electric potential 
\begin{equation}
\Phi\cong\frac{((2M)^2+Q^2)Q}{2M\sqrt{((2M)^2+Q^2)^2+(2J)^2}}. \label{ep}
\end{equation}
In the dynamical context, $E\ge M$ is not the ADM energy, but can be 
interpreted as the effective energy of the black hole, including irreducible 
mass $M$, rotational kinetic energy $\approx\frac12I\Omega^2$ and 
electrostatic energy $\approx\frac12Q^2/R$, by expanding $E\approx 
M+\frac12I\Omega^2+\frac12Q^2/R$ for $J\ll M^2$ and $Q\ll M$, where 
$J=I\Omega$ defines the moment of inertia $I\cong 
M\sqrt{((2M)^2+Q^2)^2+(2J)^2}\cong ER^2$. 

The state-space formulas 
\begin{equation}
\kappa\cong8\pi\frac{\partial E}{\partial A}\cong{1\over{4M}}\frac{\partial 
E}{\partial M}, \quad\Omega\cong\frac{\partial E}{\partial J}, 
\quad\Phi\cong\frac{\partial E}{\partial Q} \label{pd}
\end{equation}
then yield a dynamic version of the so-called first law of black-hole 
mechanics \cite{BCH}: 
\begin{equation}
L_\xi E\cong\frac{\kappa}{8\pi}L_\xi A+\Omega L_\xi J+\Phi L_\xi Q. \label{g}
\end{equation}
As desired, the state-space perturbations in the classical law for Killing 
horizons \cite{BCH}, or the version for isolated horizons \cite{IH}, have been 
replaced by the derivatives along the trapping horizon, thereby promoting it 
to a genuine dynamical law.

{\em Equilibrium.} When a growing black hole ceases to grow, the generically 
spatial trapping horizon becomes null, leading to a non-uniqueness in the 
twist. However, preservation of angular momentum suggests a natural way to 
restore uniqueness by fixing $Df\cong0$ \cite{bhd5}, thereby closing another 
gap in the paradigm. Then the dominant energy condition implies
\begin{equation}
g(\tilde\jmath,\tau)\cong g(\bar\jmath,\tau)\cong g(j,\tau)\cong0
\end{equation}
and the conserved quantities are actually preserved:
\begin{equation}
L_\xi M\cong L_\xi J\cong L_\xi Q\cong0.
\end{equation}
This indicates that local equilibrium is indeed attained when a trapping 
horizon becomes null. Then $L_\xi\kappa\cong0$, so that the surface gravity, 
which satisfies $D\kappa\cong0$ by definition (\ref{sg}), is constant where a  
trapping horizon becomes null. This is a quite general zeroth law. By the area 
law \cite{bhd}, this also shows that a black hole cannot change its angular 
momentum or charge without increasing its area.

{\em Conclusion.} Of the classical four laws of black-hole mechanics 
\cite{BCH}, the optimal form of the third law is still not clear. Generalized 
versions of the zeroth and first (\ref{g}) laws have been given above. The 
analogue of the second law is the area law derived previously \cite{bhd}, 
which can be regarded as a consequence of energy conservation and horizon type.

The new paradigm for black holes in terms of trapping horizons thereby makes 
appropriate contact with the classical paradigm, while shifting emphasis to 
more fundamental conservation laws for energy (\ref{ec}) and angular momentum 
(\ref{amc}), which include plausible contributions from gravitational 
radiation. Thus three areas in General Relativity which are intuitively 
important as physics but have been conceptually elusive, namely energy, black 
holes and gravitational radiation, appear to make sense quite generally and 
are profoundly interrelated.

Research supported by the National Natural Science Foundation of China under 
grant 10473007. Thanks to Abhay Ashtekar, Ivan Booth, Eric Gourgoulhon and 
Badri Krishnan for discussions.

\end{document}